\documentclass{article}

\usepackage{amsmath}
\usepackage{commath}
\usepackage{url}
\usepackage{cite}
\usepackage{color}
\usepackage{multirow}
\usepackage[margin=1.25in]{geometry}
\usepackage{graphicx}
\graphicspath{{./graphics/}}
\usepackage[font=small,margin=1cm]{caption}
\usepackage{pdfpages}
\usepackage{hyperref}

\usepackage{dcolumn}
\newcolumntype{C}[1]{>{\centering\arraybackslash}p{#1}}

\begin{document}
\begin{center}
\section*{Executive Summary}
Proposal to PAC51: Color Transparency in Maximal Rescattering Kinematics
\end{center}
\subsection*{Main physics goals}
The physics goal of this experiment is to observe the QCD prediction of color transparency in protons. If observed, this would be the  discovery of color-neutral protons fluctuating into small transverse size configurations within nuclei as predicted by $\rm SU(3)_C$-based  QCD. This bridges our descriptions of nuclear physics and fundamental QCD theory. The momentum regime at which the onset of color transparency is observed tells us where factorization theorems become relevant. Previous attempts to observe the onset of color transparency in protons have been unsuccessful. The choice of nuclei and kinematics in these attempts were susceptible to rapid expansion effects of the point-like hadron. Therefore, this experiment is designed to study the d$(e,e'p)n$ reaction in kinematics where the expansion effects are minimal (i.e., the distance between the hard interaction point and the interaction with the second nucleon is small). In these kinematics, the deuteron's wave function is well-known and calculable (as contrasted with heavier nuclei). Specifically, the deuteron rescattering kinematics are well known, the distances involved in rescattering are less than 1~fm, and the dynamics are well described by the generalized eikonal approximation. This experiment will exploit the well-understood and calculable dynamics of rescattering effects in the deuteron, which include minimizing effects from the expansion of the hadron, to make the best experimental probe possible for the observation of hadrons in a point-like configuration.

\subsection*{The proposed measurements and observables}
This experiment will measure the d$(e,e'p)n$ reaction in kinematics where rescattering dominates final state interactions. The observable will be the cross section ratio of the high-to-low missing momentum for various data points in $Q^2$ such that the high missing momentum regime is dominated by double scattering as compared to the low missing momentum regime. A significant decrease in this cross section ratio with increasing $Q^2$ is consistent with predictions for the onset of color transparency in protons. 

\subsection*{Specific requirements on detectors, targets, and beam}
This experiment will use the standard Hall C configuration with both the SHMS and HMS spectrometers in coincidence. To probe the highest kinematic points with reasonable beam time, we require an incident 11 GeV electron beam on a 25~cm long liquid deuterium target which is possible after the ESR-2 upgrade.

\subsection*{Submission}
This proposal is a first time submission to the PAC and is a follow-up from the Letter of Intent submitted to PAC50 (LOI12-22-001).

\title{Proposal to PAC 51:\\ Color Transparency  in Maximal Rescattering Kinematics}
\newpage

\author{
S.~Covrig Dusa, M.~Diefenthaler, D.~Gaskell, D.W.~Higinbotham~(co-spokesperson),\\
M.K.~Jones, D.~Nguyen, H.~Szumila-Vance~(co-spokesperson\footnote{Contact Person: \texttt{hszumila@jlab.org}}), A.~Tadepalli\\
Thomas Jefferson National Accelerator Facility,\\ Newport News, Virginia 23606, USA\\
\and
J.~Arrington, T.~Hague, S.~Li~(co-spokesperson), J. Rittenhouse~West~(co-spokesperson)\\
Lawrence Berkeley National Laboratory,\\ 
Berkeley, California 94720, USA\\
\and
C.~Fogler, L.~Weinstein, C.~Yero~(co-spokesperson)\\
Old Dominion University, Norfolk, VA 23529, USA\\
\and
W.~Boeglin, P.~Markowitz, M.~Sargsian, W.~Cosyn\\
Florida International University, Miami, FL 33199, USA\\
\and
H.~Bhatt, B.~Devkota, D.~Dutta, A.~Nadeeshani, B.~Tamang, E.~Wrightson\\
Mississippi State University, Mississippi State, MS 39762, USA\\
\and
M.~Strikman\\
Pennsylvania State University, State College, PA 16801, USA\\
\and
N.~Fomin\\
University of Tennessee, Knoxville, TN 37996, USA\\
\and
G.~Huber\\
University of Regina, Regina, SK S4S0A2, Canada\\ 
\and
M.~Elaasar\\
Southern University at New Orleans, New Orleans, LA 70126, USA\\
\and
D.~Androi\'{c}\\
University of Zagreb, Zagreb, Croatia\\
\and
W.~Li\\
Center for Frontiers in Nuclear Science, Stony Brook, 11794, NY, USA\\
Stony Brook University, Stony Brook, 11794, NY, USA\\
\and
G.~Niculescu\\
James Madison University, Harrisonburg, VA 22807, USA\\
\and
C.~Ayerbe Gayoso\\
William and Mary, Williamsburg, VA 23187, USA\\
\and
N.~Santiesteban\\
University of New Hampshire, Durham, NH 03824, USA\\
\and
E.~Piasetzky\\
Tel Aviv University, Tel Aviv 6997801, Israel\\
}
\date{\vspace{-5ex}}
\maketitle
\begin{figure}[!h]
\centering
\includegraphics[width=0.3\textwidth]{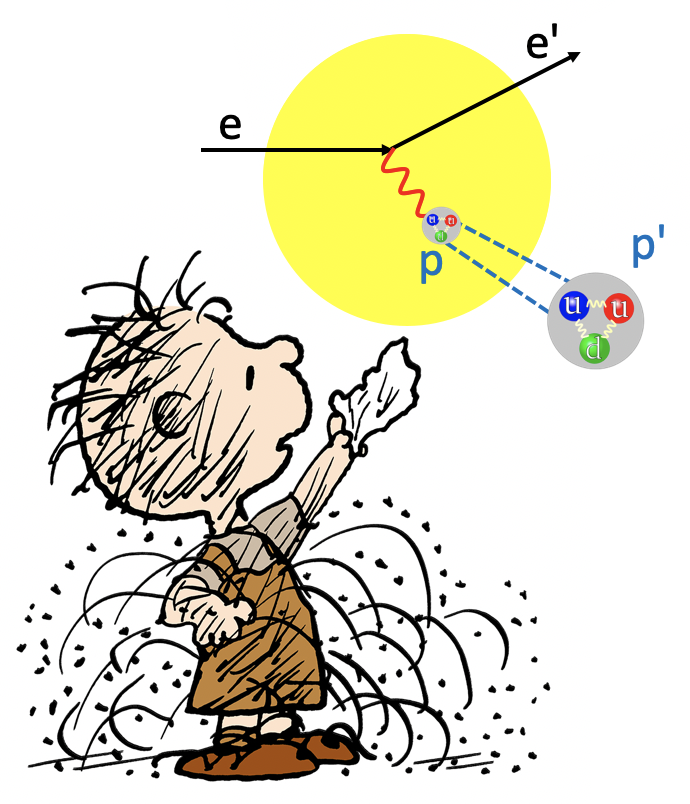}
\caption*{\textit{Color Transparency in ``Dirty'' Kinematics}}
\end{figure}
\begin{abstract}
With the current highest beam energy at Jefferson Lab and traditional methods, we have exhausted our sensitivity for observing the onset of proton color transparency in a nucleus in $A(e,e^{\prime} p)$ parallel scattering kinematics for up to $Q^2=14$~GeV$^2$. One of the disadvantages in $A(e,e{^\prime} p)$ experiments is that even if 
a point-like color singlet is produced at such $Q^2$, its expansion is unconstrained over the  full radius of the nuclei, with the potential to significantly reduce the 
size of the color transparency effect. Therefore, in order to be sensitive to the effects of color transparency, we enhance the sensitivity of the measurement to the production of a point-like color neutral object prior to the onset of wavefunction expansion.

In this experiment, we propose a color transparency measurement in  maximal rescattering (``dirty'') kinematics in deuterium where final-state interactions (FSIs) are known to be huge effects, thereby enhancing our sensitivity to a reduction in FSIs indicative of color transparency. The kinematics in exclusive processes in deuterium can be precisely chosen such that the inter-nucleon
distances of the struck and spectator nucleon lead to well-controlled FSIs. In the high $Q^2$ regime expected with the onset of the CT, these reactions enable observations of the
formation of the point-like configuration (PLC) before expansion. By measuring in these kinematics as a function of 
increasing $Q^2$, we demonstrate both an enhanced sensitivity to the onset of color transparency as well as observing the PLC before any expansion effects reduce its signal.

We propose to measure the d$(e,e'p)n$ proton knockout cross sections at different neutron recoil momenta, for kinematics with known large FSIs,  to extract the proton nuclear transparency as a function of increasing $Q^2$. This can be achieved by measuring the transparency ratio of the detected protons 
at two different recoil momenta for the neutrons which  
are characterized by different FSI effects. We will measure the transparency over the range of $Q^2=8$--$15$~(GeV/$c)^2$. We request a total of 95~days of beam time: 91~days of physics measurements and 4~days of calibrations, configuration changes, and background measurements.

\end{abstract}

\section{Introduction}
\label{sec:intro}
The strong force is well-described at low energies and long distances by the exchange of color-neutral mesons between nucleons while at high energies and short distances a perturbative quantum chromodynamics (QCD) description with color-charged quark and gluonic degrees of freedom is valid. The boundary between these two regimes is a mystery, but each of them enables precise descriptions of nuclear interactions in their respective domains. While QCD is the theory of the strong interaction and the basis for nuclear phenomena, there are many missing links needed for it to directly describe matter in terms of nucleons and nuclei.  The onset regime for color transparency in the proton, if discovered, is a direct link between QCD and nuclear physics.

\subsection{Color Transparency}
\label{sec:ct-intro}
For exclusive processes at high momentum transfer, QCD predicts the phenomenon of color transparency (CT)\cite{Brodsky:1988xz,Brodsky:1994kf,Dutta:2012ii,Brodsky:2021dze} in which one can preferentially scatter off hadrons that have fluctuated into a point-like configuration (PLC).  The quarks within the PLC have formed an object of highly reduced transverse size which enables it to exit the nucleus with no further interactions by virtue of the $\rm SU(3)_C$-based QCD prediction of color-anticolor neutrality for mesons and tri-color neutrality for baryons.  Due to the point-like nature of the PLC, the external color fields of the hadron cancel (the object is color-neutral), thereby suppressing further gluonic interactions with the nuclear medium.  The PLC is maintained for some period of time described by the expansion factor $\tau$ (expansion time in the PLC rest frame). CT is uniquely predicted by QCD and exists at high momentum transfer. The momentum regime of the onset of CT is of great interest because it sits at the transition between the low energy hadronic description and the high energy partonic description of nuclei. 

The transparency variable, $T$, is defined as the cross section per nucleon for a process on a bound nucleon relative to that of a free nucleon such that $T=\sigma_A/(A\sigma_0)$. The onset of color transparency is experimentally measured as an increase in the transparency with increasing momentum transfer (or momentum transfer squared, $Q^2$, as is commonly used).  The subfield of QCD known as holographic light-front QCD (HQCD) \cite{Brodsky:2014yha,Dosch:2015nwa}, which is a set of tools for studying the dynamics of strongly coupled quantum field theories such as QCD, has predicted a two-stage onset of CT for the proton and one stage onset for the neutron.  The onset of CT for the proton is predicted at $14$ GeV$^2$ for its first stage, while the onset for the neutron and second stage proton CT occurs at $22$ GeV$^2$ \cite{Brodsky:2022bum} (discussed in Sec.~\ref{sec:ct-intermediate}).  The experiment described herein should be sensitive to the onset of CT in the proton with these predictions.

CT is relevant to QCD factorization theorems and the onset ($Q^2$) of CT is where leading order perturbative QCD is applicable. Factorization theorems derived for deep inelastic exclusive processes are required for accessing Generalized Parton Distributions (GPDs)~\cite{Brodsky:1994kf,Collins:1996fb,Frankfurt:1999fp,Diehl:1998dk}. GPDs map the nucleon wave function and describe the transverse momentum and angular momentum carried by quarks in the proton~\cite{Ji:1996nm,Radyushkin:1997ki}. The full mapping of GPDs is the subject of significant experimental efforts in nuclear physics today. Furthermore, the suppression of final state interactions is essential to Bjorken scaling in the deep-inelastic regime at small $x_B$~\cite{Frankfurt:1988nt}.

An important detail of CT is that the PLC will expand as it moves through the nucleus because it is not an eigenstate of the Hamiltonian. The expansion time, $\tau$ (or, equivalently, the coherence length $l_c$ 
over which the hadron travels as it evolves from a PLC to its normal size) 
is unknown, but there are several estimation methods in the literature.  A widely used estimate gives $\tau \approx (E_h / m_h) \tau_{0}$, where $E_h$ is the 
laboratory  
energy, $m_h$ is the mass of the ground state hadron, 
and $\tau_{0} \approx 1 ~\mathrm{fm}$ is the characteristic hadron rest frame time~\cite{Frankfurt:1992zp}, which shows that an increase in $E_h$ allows a PLC to remain small for a longer time, thus increasing the suppression of FSIs. 

 For the knocked-out proton in the kinematics of the recent experiment~\cite{PhysRevLett.126.082301}, i.e., for $Q^2$ = 8--14 GeV$^2$ on a carbon nucleus, holographic light-front QCD techniques give a coherence length of $l_c \sim $ 2--3 fm~\cite{Caplow-Munro:2021xwi}. This is more than sufficient to observe CT of the proton in deuterium, but we note that there are a few differences in analysis from the recent HQCD  
paper \cite{Brodsky:2022bum}.  The naive parton model in which PLC constituents separate at the speed of light gives a transverse hadron 
size of $x_{\tau} \sim \tau \sim(E_h / m_h)^{-1} z$, where $z$ is the longitudinal coordinate in the lab frame and $E_h/m_h$ is the time dilation factor as before. This gives a coherence length estimate of $l_{c} \approx E_h / m_{h} ~\mathrm{fm}$ for the PLC~\cite{Farrar:1988me}. The uncertainty principle gives a maximum coherence length, $l_{c} \sim \frac{1}{\Delta M} \frac{p'}{m_{h}}$, where $\Delta M$ is a characteristic excitation energy and $p'$ is the 3-momentum of the particle in the intermediate excited state~\cite{Sargsian.2003}.  The quantum diffusion model is inspired by perturbative QCD and gives a coherence length $l_{c} \simeq\left\langle 1 /\left(E_{\rm ex}-E_{h}\right)\right\rangle \simeq 2 p'\left\langle 1 /\left(m_{\rm ex}^{2}-m_{h}^{2}\right)\right\rangle$, where $m_\text{ex}$ is the mass of the hadron in the excited (i.e., intermediate) state~\cite{Farrar:1988me}. This can be seen by considering the intermediate state propagator: $l_c = {1\over E_\text{ex}-E_h}$, with $E_h = \sqrt{m_h^2+p^2}$ and $E_\text{ex} = \sqrt{m^2_\text{ex}+p'^2}$. For high-energy kinematics, $p'\gg m_h, m_\text{ex}$, the coherence length becomes 
$l_c\approx {2p'/\Delta M^2}$,
where $\Delta M^2 = m^2_\text{ex}- m^2_h$.  The smaller size of deuterium assists in minimizing the expansion effects by minimizing the PLC travel length prior to collision site exit.

Previous estimates extracted $\tau$ from the BNL $(p,pp)$ reactions assuming that the transparency rise was from CT. The same value of $\tau$ was used to predict CT in $(e,e'\pi)$, $(e,e'\rho)$, and $(e,e'p)$. The $\tau$ predictions are consistent with the meson production data but have been unsuccessful in predicting the onset of CT in protons. There is no strict reason that the expansion time should be the same for meson and baryons. If protons have a smaller $\tau$ than mesons, the onset of CT is delayed to higher outgoing proton energy (proportional to $Q^2$). At $Q^2=20$~GeV/c$^2$, $\tau=11$ fm$^{-1}$ ($1/\tau\approx 3\times$~radius of $^{12}$C). This indicates that up to $Q^2=20$~GeV/c$^2$, there is motivation to measure the onset of CT in baryons to determine if PLC formation could be related to the observed EMC Effect~\cite{miller_email}. In such models, the EMC effect is explained as a result of the suppression of PLCs in the bound nucleon wave function, which is emphasized when the bound nucleon is highly-virtual (states that are also dominated by short range correlated nucleon pairs)~\cite{Hen:2016kwk}.

In deuterium, the kinematics in exclusive processes can be precisely chosen such that the inter-nucleon
distances between the struck and spectator nucleon lead to well-controlled FSIs. In the case
of high $Q^2$, with the onset of the CT regime, these reactions enable observations of the
formation of the PLC before expansion and control its expansion in the FSI process.

\subsection{Previous CT measurements at intermediate energies}

Measurements in the intermediate energy regime are of direct interest to the observation of the onset of CT. The onset of CT is favored to be observed at lower energy in mesons than baryons since only two quarks must come close together, and the quark-antiquark pair is more likely to form a PLC~\cite{Blaettel:1993rd}. Pion production measurements at JLab reported evidence for the onset of CT \cite{Clasie:2007aa} in the process $e + A \rightarrow e + \pi^+ + A^{*}$. The results of the pion electroproduction experiment showed that both the energy and $A$ dependence of the nuclear transparency deviate from conventional nuclear physics and are consistent with models that include CT. The pion results indicate that the energy scale for the onset of CT in mesons is $\sim 1$~GeV.

A CLAS experiment studied $\rho$-meson production from nuclei, and the results also indicated an early onset of CT in mesons~\cite{ElFassi:2012nr}. Previous $\rho^0$ production experiments had shown that nuclear transparency also depends on the coherence length, $l_c$, which is the length scale over which the $q\bar{q}$ states of mass $M_{q\bar{q}}$ can propagate. The CLAS collaboration further measured the transparency for incoherent exclusive $\rho^0$ electroproduction off carbon and iron relative to deuterium~\cite{ElFassi:2012nr} using a 5~GeV electron beam. An increase of the transparency with $Q^2$ for both C and Fe was observed. The rise in transparency was found to be consistent with predictions of CT by models \cite{Frankfurt:2008pz,Gallmeister:2010wn} which had accounted for the increase in transparency for pion electroproduction. The $\pi$ and $\rho$ electroproduction data set the energy range to be at a few GeV for the onset of CT in mesons. 

While several experiments have observed a rise in the transparency for mesons, the observation of the onset of color transparency in baryons remains ambiguous. The focus of this proposal is to search for the onset of CT in baryons using deuterium where the kinematics can be optimized to enhance the signal and decrease susceptibility to expansion effects.

\subsection{CT in baryons at intermediate energies}
\label{sec:ct-intermediate}
The observed signal of the onset of CT in baryons remains ambiguous. The first attempt measured large angle $A(p,pp)$ scattering at Brookhaven National Lab (BNL)~\cite{Carroll:1988rp,Mardor:1998zf,Leksanov:2001ui,Aclander:2004zm}. These experiments measured the transparency as the quasi-elastic cross section from the nuclear target to the free $pp$ elastic cross section. The results of these experiments indicated a rise the in the transparency for outgoing proton momenta of 6--9.5~GeV/c consistent with models of CT. However, the transparency was observed to decrease at higher momenta between 9.5--14.4~GeV$/c$. This decrease is inconsistent with CT as a plateau after the onset as predicted for CT. The results of these experiments are not fully explained by anyone. The BNL experiments have an additional complication in that the incident proton suffers a reduction of flux in medium and must be included in any transparency calculation. Possible explanations include an elastic energy dependent cross section due to nuclear filtering from the Landshoff mechanism~\cite{Kundu:1998ti,Ralston:1990jj} or excitation of charm resonances beyond the charm production threshold~\cite{PhysRevLett.64.1011}. 

\begin{figure}[htb]
\centering
\includegraphics[width=0.7\textwidth]{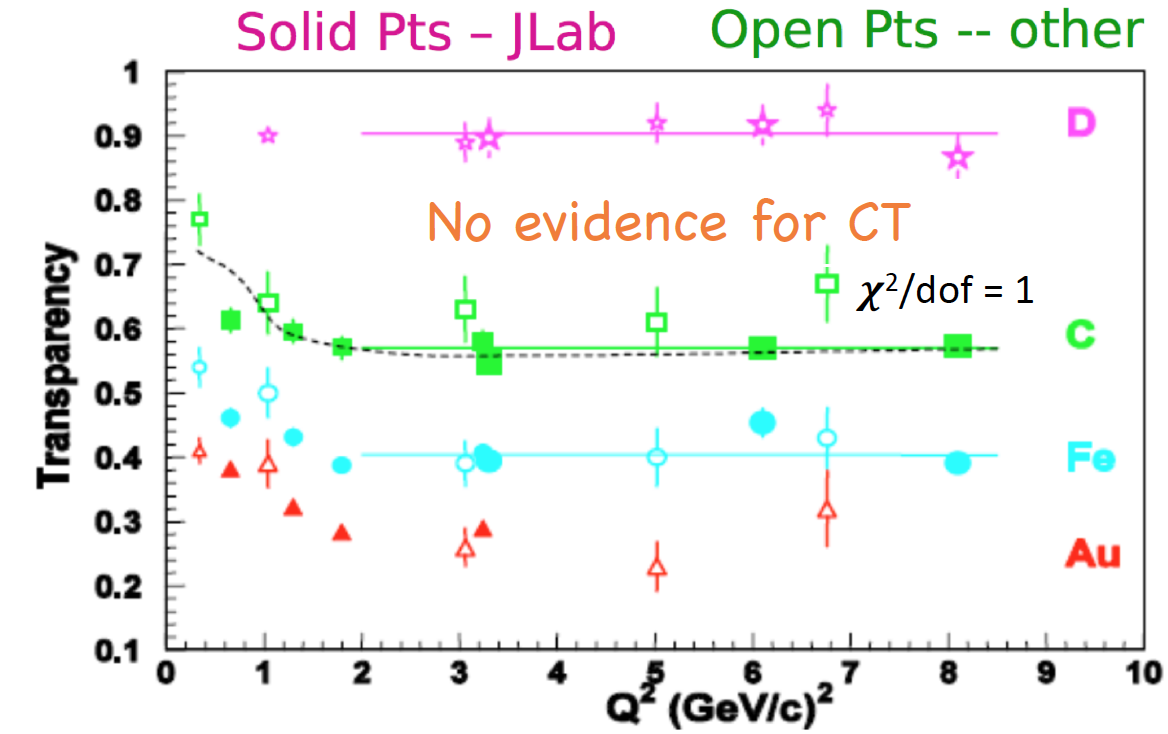}
\caption{\label{fig:baryonsCT} The combined results are shown for $A(e,e'p)$ searches for CT at MIT-Bates, SLAC, and JLab. A fit is shown that indicates no significant rise consistent with CT.~\cite{Abbott:1997bc,Makins:1994mm,Garrow:2001di}.} 
\end{figure}

The $(e,e'p)$ reaction is simpler to understand than the $(p,pp)$ reaction as only the final state proton suffers a reduction of flux and needs to be considered in the measurement. The first experiments using an electron beam to measure CT were at SLAC~\cite{Makins:1994mm,ONeill:1994znv} followed by experiments at JLab~\cite{Abbott:1997bc,Garrow:2001di}. In high $Q^2$ quasielastic $(e,e'p)$ scattering from nuclei, the electron scatters from a single proton, which has some associated Fermi motion \cite{Frullani}. In the plane wave impulse approximation (PWIA), the proton is ejected without final state interactions with the residual $A-1$ nucleons. The measured $A(e,e'p)$ cross section would be reduced compared to the PWIA prediction in the presence of final state interactions, where the proton can scatter both elastically and inelastically from the surrounding nucleons as it exits the nucleus. The deviations from the PWIA model measure the nuclear transparency. In complete CT, the final state interactions would vanish, and the nuclear transparency would plateau. This is in contrast to the conventional picture where the nuclear transparency is expected to follow the same energy dependence as the relatively energy-independent $NN$ cross section.

The combined results of searches for CT in the $A(e,e'p)$ reaction are shown in Fig.~\ref{fig:baryonsCT}. The results indicate that there is not significant rise up to $Q^2=8$~GeV/c$^2$. If translating the $A(p,pp)$ BNL reaction by relating $t$ to the electron-scattering $Q^2$, then the effect observed at BNL has already been ruled out by the experimental results shown in Fig.~\ref{fig:baryonsCT}. The most recent result on this topic was performed in a recent Hall C 12~GeV experiment~\cite{PhysRevLett.126.082301} which sought to eliminate the possibility that the increase in transparency observed at BNL is due to the higher incident proton momentum than was observed at JLab during the 6~GeV era. The onset of CT depends both on momentum and energy transfers, affecting the squeezing and freezing, respectively. Since $A(e,e'p)$ scattering measurements are carried out at $x_B=1$ kinematics, they are characterized by lower energy transfers compared to the momentum transfer. It is possible that the 6~GeV era experiments were unable to satisfy the energy-dependent freezing requirement. 
\begin{figure}[htpb]
\centering
\includegraphics[width=0.7\textwidth]{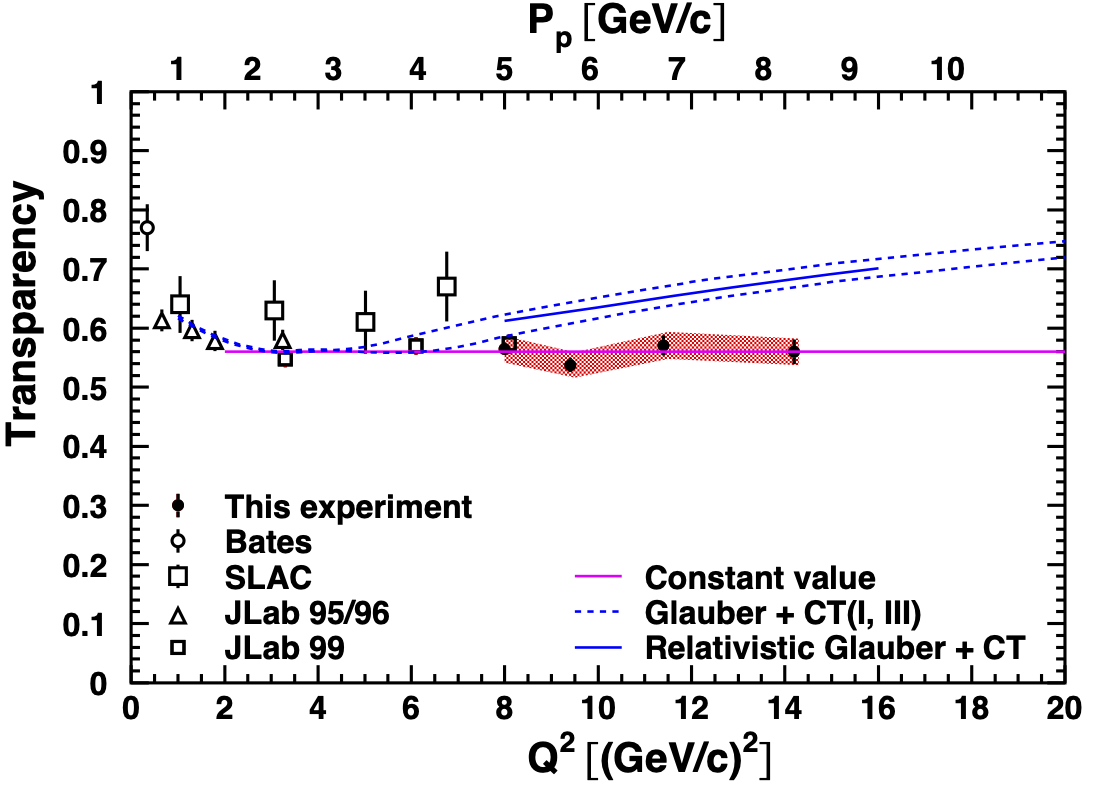}
\caption{\label{fig:cctplot} All present carbon electron-scattering results are shown including the latest Hall C measurement~\cite{PhysRevLett.126.082301} indicating no onset in CT up to $Q^2=14.2$~GeV$^2/c^2$.} 
\end{figure}
The results from the most recent experiment in Hall C~\cite{PhysRevLett.126.082301} are shown in Fig.~\ref{fig:cctplot}, but they do not indicate a rise in CT up to $Q^2=14.2$~GeV/c$^2$ within a 6\% uncertainty. It should be noted that these recent results searched for the onset of CT in parallel kinematics (such that the incident struck proton's momentum is parallel to the electron momentum transfer, $\vec{q}$). The BNL measurements were taken with perpendicular kinematics which in electron scattering is generally dominated by nucleon re-scattering~\cite{Barbieri:2004nf,E97-006:2005jlg,Rohe:2004dz}. The parallel kinematics in electron scattering are characterized as a regime of already low FSIs and search for a reduction of FSIs. While the previous proton knockout experiments spanning $Q^2$ did not see a signal consistent with CT, they were all taken in parallel kinematics with already low initial FSI effects, and they were unable to separate the production mechanism from the expansion of the PLC. 

The recent proton results of ~\cite{PhysRevLett.126.082301} reinvigorated the physics community and has since prompted significant ongoing discussion and a re-examination of the CT models. A workshop dedicated to these efforts was held, titled \href{https://indico.jlab.org/event/437/contributions/}{The Future of Color Transparency and Hadronization Studies at Jefferson Lab and Beyond}. Several key ideas have since risen out of this workshop (see a few select topics: \cite{Brodsky:2022bum,Miller.2022,Jain.Pire.Ralston.2022,Huber.2022,gallmeister.mosel.2022}). 

Specifically, \cite{Caplow-Munro:2021xwi} attributes the lack of observation of the CT onset in the Hall C data to the Feynman Mechanism. In the Feynman Mechanism, the virtual photon interacts with a single quark that carries a large momentum fraction of the proton such that the proton's total quark configuration was never in a PLC. As briefly described in Sec.~\ref{sec:ct-intro}, HQCD predictions demonstrate that the onset of CT in protons is expected to occur in two stages where the initial onset $Q^2$ for the spin-conserving (twist-3) Dirac form factor occurs just above 14~(GeV/$c)^2$. The later onset for the spin-flip Pauli (twist-4) form factor occurs at $Q^2>22$~(GeV$/c)^2$ \cite{Brodsky:2022bum}. The theoretical work predicting either the onset of CT in the proton at momentum transfer $Q^2>14~\rm GeV^2$ \cite{Brodsky:2022bum} or no PLC formation from the Feynman mechanism \cite{Caplow-Munro:2021xwi} is observable through the cross section dependence on the separation of proton constituents and whether it falls or not with increasing momentum transfer. In the Feynman mechanism vs. CT work and earlier work \cite{Frankfurt:1993es}, the cross section dependence on the square of the transverse separation distance between proton constituents ($b^{2}=\sum_{i<j}\left(b_{i}-b_{j}\right)^{2}$) is given by $\lim_{b \rightarrow 0} \sigma\left(b^{2}\right) \propto b^{2}$. If $b^2(Q^2)$ is shown to decrease as $Q^2$ increases, meaning if the cross section decreases as $Q^2$ increases and therefore the transparency $T$ increases, a PLC is said to be possible/admitted. While HQCD CT work uses a different variable to define the approach to PLC, $\mathbf{a}_{\perp}=\sum_{j=1}^{n-1} x_{j} \mathbf{b}_{\perp j}$ as the size of the PLC, they are related by a weighting of $x_B$ \cite{Brodsky:2022bum}. For fractional $x_B$, this decreases the cross sectional dependence as compared to \cite{Caplow-Munro:2021xwi} but the overall effect is the same: CT in the proton predicts a decrease in FSI while the Feynman mechanism does not.

The biggest challenges in exploring CT in ($e,e^\prime,N$) experiments in nuclei are twofold:
{\em First}, in these processes the focus is on the effects of absorption of a produced baryon (proton),  
which (neglecting expansion effects) is proportional to $\sim$ $(1- {\sigma_{PLC,N}(Q^2)}\over \sigma_{NN}$, where $\sigma_{PLC,N}$ and $\sigma_{NN}$ are total  cross sections of PLC-nucleon and nucleon-nucleon scattering;
{\em Second}, an inability to control expansion 
effects at intermediate energies.  The problem in this case is that the PLC-to-baryon formation 
length includes the full radius of the nuclei which is of order a few Fermis.

The key idea presented in this proposal is to explore color transparency in kinematics in which the scattering cross section is 
defined exclusively by the final state rescattering of a PLC off the spectator nucleon. This addresses both 
of the problems stated above; {\em First}, the cross section of the process (neglecting expansion effects) is proportional to 
$\sim {\sigma^2_{PLC,N}(Q^2)\over \sigma^2_{NN}}$; {\em Second}, the expansion effect is confined to the distance between
PLC production and the spectator nucleon. Most importantly, if one can control 
the momentum of the rescattered nucleon one can control the distances over which expansion took place.  Therefore, using a deuteron target in which the momentum of the spectator nucleon is controlled 
one can achieve maximal sensitivity to CT effects.

 While the idea of exploring CT in rescattering kinematics was proposed prior to the 12~GeV upgrade~\cite{Frankfurt:1994kt, Sargsian.2003}, and the first such 
 measurements were made in Hall B\cite{PhysRevLett.98.262502} for 6~GeV energies, several arguments make the current proposal timely and impactful in our opinion.
 The Hall B measurements included the fact that the CLAS measurement\cite{PhysRevLett.98.262502}
had to integrate the cross section over the  rather wide kinematical ranges and 
detector effects had to be included in the calculations to compare 
with the data.  In addition, higher accuracy measurements in Hall A came several 
years later which allowed a quantitative assessment of the accuracy of theoretical 
calculations performed by several 
groups\cite{LAGET200549,Orden_2014,Sargsian:2009hf}.

Finally, the current updated predictions for CT effects  include the constraints from the recent Hall C proton measurement indicating that there is parameter space to observe the signal of CT at $Q^2>12$~(GeV$/c)^2$ as described in \cite{physics4040092}. 

\section{\texorpdfstring{d$(e,e'p)n$}{d(e,e'p)}}

We propose in this experiment to search for the onset signal of CT using deuterium. Deuterium is the simplest target nucleus, and the wave function in collisions is well-described using the generalized eikonal approximation \cite{Capel:2019zor,Sargsian:2004tz}. The kinematics can be precisely chosen such that the inter-nucleon distances of the struck and spectator nucleon lead to well-controlled FSIs~\cite{Frankfurt:1994kt} and subsequently, enables observations of the formation of the PLC before expansion. The proton knockout reaction in deuterium, d$(e,e'p)$ is described as:
\begin{equation}
    e+d\rightarrow e'+p+n
\end{equation}
Deuterium is a lighter nucleus compared to the target nuclei used in recent CT experiments and should therefore minimize expansion effects of any observed PLC.

\begin{figure}[htb]
\centering
\includegraphics[width=0.7\textwidth]{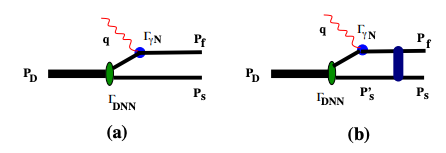}
\caption{\label{fig:feynmanRescatter} Electron scattering interaction probing deuterium with a) breakup and b) rescattering effects~\cite{SargsianTalk.2021}.} 
\end{figure}

While the inter-nucleon separation is relatively large in deuterium which minimizes deviations with respect to Glauber and should detract from the usefulness of deuterium as a target for CT, there exists an interference between the Born term and the rescattering amplitude of the cross section. The calculation of the cross section shows an increase when the rescattering is included (the square of the rescattering amplitude is dubbed the ``double scattering" term)~\cite{Frankfurt:1994kt}, see rescattering in Fig.~\ref{fig:feynmanRescatter}b. 

\begin{figure}[htb]
\centering
\includegraphics[width=0.7\textwidth]{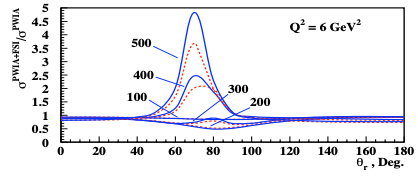}
\caption{\label{fig:rescatterCalc} The dependence of the cross section ratio taken with respect to the PWIA on the recoil angle of the neutron ($\theta_r$) for different values of the recoil neutron momenta is shown for $Q^2=6$~(GeV$/c)^2$~\cite{Sargsian:2009hf}.} 
\end{figure}

These rescattering effects decrease in the presence of a PLC. In this way, the regions of high FSIs in deuterium may be compared to kinematics with low FSIs with a high degree of sensitivity to PLC formation so that a deviation may be observed as a function of increasing $Q^2$. The kinematics of double scattering reactions is optimized where the undetected partner recoil nucleon has a large perpendicular (with respect to $\vec{q}$) momentum (i.e. $p_{\perp}\geq200$~MeV$/c$ for the recoil nucleon). See Fig.~\ref{fig:rescatterCalc}.   

\begin{figure}[htb]
\centering
\includegraphics[width=0.5\textwidth]{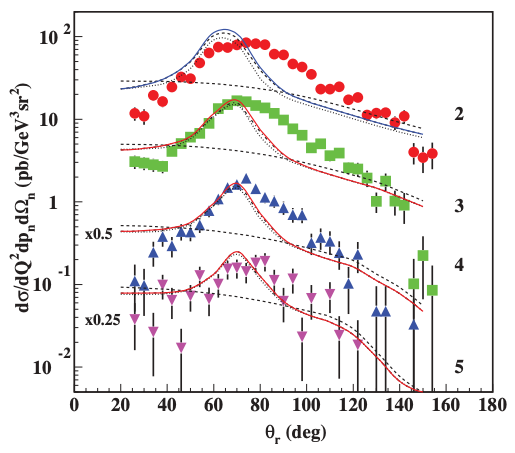}
\caption{\label{fig:recoilXShigh} Data from \cite{PhysRevLett.98.262502} shows the dependence of the differential cross section on the recoil neutron angle and momentum where the momentum of the recoil neutron for these data is restricted to be between 400--600~MeV$/c$. The dashed line shows the PWIA calculation while the other lines show calculations that include PWIA and other FSI effects~\cite{Sargsian:2009hf}. The different colors of data points correspond to different $Q^2$ ranging from 2--5~(GeV$/c)^2$, increasing from top to bottom. 
The detector effects are not included in the theoretical comparison.} 
\end{figure}

\begin{figure}[htb]
\centering
\includegraphics[width=0.5\textwidth]{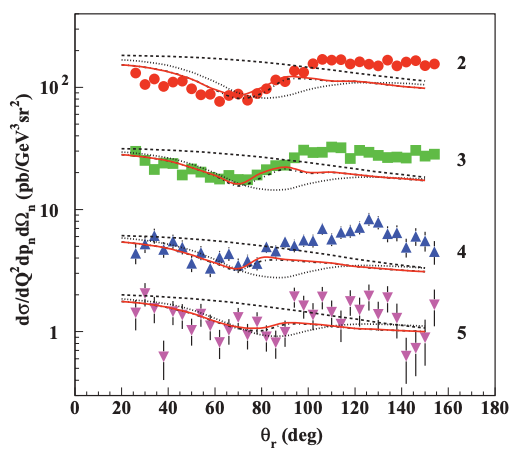}
\caption{\label{fig:recoilXSlow} Data from \cite{PhysRevLett.98.262502} shows the dependence of the differential cross section on the recoil neutron angle and momentum where the momentum of the recoil neutron for these data is restricted to be between 200--300~MeV$/c$. The dashed line shows the PWIA calculation while the other lines show calculations that include PWIA and other FSI effects~\cite{Sargsian:2009hf}. The different colors of data points correspond to different $Q^2$ ranging from 2--5~(GeV$/c)^2$, increasing from top to bottom.} 
\end{figure}

\begin{figure}[htb]
\centering
\includegraphics[width=0.7\textwidth]{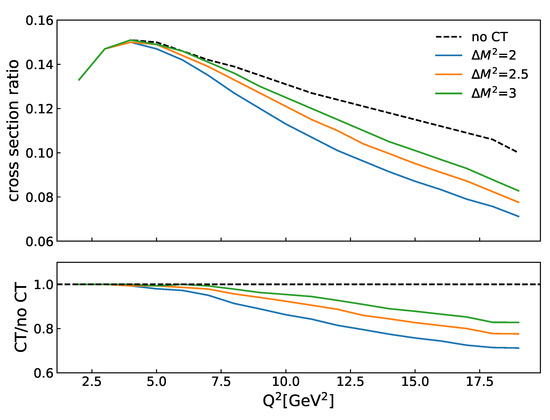}
\caption{\label{fig:rescatteringPrediction} Predictions of cross section ratio $R=\frac{\sigma(p_r=400\textrm{MeV/c})}{\sigma(p_r=200\textrm{MeV/c})}$ in the d$(e,e'p)n$ reaction from~\cite{SargsianTalk.2021}. The solid curves show calculations of the CT effects with the expansion parameter $\Delta M^2$=2, 2.5, and 3 GeV$^2$~\cite{physics4040092}. %
The red band shows the updated calculation corresponding to the new CT predictions using the constraints and uncertainties of the recent Hall C results. The deviation of this band with respect to the dashed black line for the ``No CT" prediction indicates the $Q^2$ region of interest. The ``No CT" prediction is calculated using the generalized eikonal approximation. } 
\end{figure}

We will measure the ratios of proton knockout from deuterium where PLC effects are anticipated to be high and compare them to kinematics where the PLC effects are expected to be lower as a function of increasing momentum transfer, $Q^2$. Double scattering accesses inter-nucleons distances on the order of 1-2~fm. Access to distances of this small magnitude will enable us to observe PLCs and to help constrain the PLC expansion rate which is anticipated to be rather large due to the lack of recent experimental observation~\cite{Sargsian.2003}. The larger the momentum of the spectator nucleon, the smaller the inter-nucleon distance and thus, the shorter the distance between the production and rescattering vertices leading to higher FSI effects. In specific kinematics, the ratio of the measured cross sections between high momentum spectator nucleons and low momentum spectator nucleons, $R$, is particularly sensitive to the effects of CT because the numerator, in a regime of high FSIs, will decrease with CT due to an overall reduction in rescattering whereas the denominator, characterized by low FSIs, will increase. Here we define the missing momentum, $p_\text{miss}$ as the momentum of the undetected spectator nucleon:
\begin{equation}
R(Q^2)=\dfrac{\sigma(p_\text{miss}~\text{large}; Q^2)\downarrow}{\sigma(p_\text{miss}~\text{small}; Q^2)\uparrow}
\end{equation}
\noindent Here, $R$ is expected to decrease with increasing $Q^2$ in the presence of PLCs, the arrows at the tail end of the numerator and denominator denote their relative change after including CT. The regime where $p_\text{miss}$ is large is chosen in the range of $300-600$~MeV$/c$ where double scattering dominates. Also, the reconstructed angle of the recoiling nucleon should be in the range of 60--90$^{\circ}$ where FSI effects deviate most from PWIA calculation. The increase in the cross section is shown in data from \cite{PhysRevLett.98.262502} in Fig.~\ref{fig:recoilXShigh}.  
In the regime where $p_\text{miss}$ is small, we are interested in nucleons with momenta $<200$~MeV$/c$ where Glauber screening is expected to play a large effect. Previous data from \cite{PhysRevLett.98.262502} shows the measured cross sections for the lower momenta recoiling neutrons in Fig.~\ref{fig:recoilXSlow}.
Our ratio will be taken with respect to data and will be compared to ratios constructed from theory. Note that this approach is unique from previous electron scattering proton knockout CT experiments which compare data directly to theoretical calculations. Our proposed experimental reach is shown in Fig.~\ref{fig:rescatteringPrediction}. 

Here, the transparency ratio on the vertical axis is taken to be the ratio for high ($p_r=400$~MeV$/c$) and low ($p_r=200$~MeV$/c$) recoiling neutron momenta in the d$(e,e'p)n$ reaction. By taking the ratio of the cross sections for the high and low recoiling nucleons, we see that as we scatter from a PLC (and hence experience less FSI effects), the observed cross section for the high momentum recoiling nucleons decreases as the observed cross section for the low momentum recoiling nucleons increases, and we observe a deviation from traditional Glauber calculations. In Fig.~\ref{fig:rescatteringPrediction}, the new prediction to observe such an effect is shown to include the uncertainties and constraints from the recent Hall C measurement. A relative $Q^2$ dependence of the FSIs could indicate a regime for the onset of CT. This proposal will focus on the use of d$(e,e'p)n$ as a tool to explore CT in kinematics that were elusive in the previous proton experiments. 

The signal in both the regions of low and high $p_\text{miss}$ is optimized where the light cone momentum fraction of the nucleus carried by the recoil nucleon, $\alpha$, is approximately 1 (i.e. $\alpha$ is defined as $\alpha=(E_n-p_n\text{cos}\theta_{\gamma n})/m_n\rightarrow 1$ in terms of the final state spectator nucleon energy, momentum, mass, and angle). The model we use to study the deuteron and make such comparisons is from AV18 for the nucleon momentum distribution~\cite{PhysRevC.89.024305}. We further make comparisons in terms of the PWIA and the FSIs. 

\begin{figure}[htb]
\centering
\includegraphics[width=0.4\textwidth]{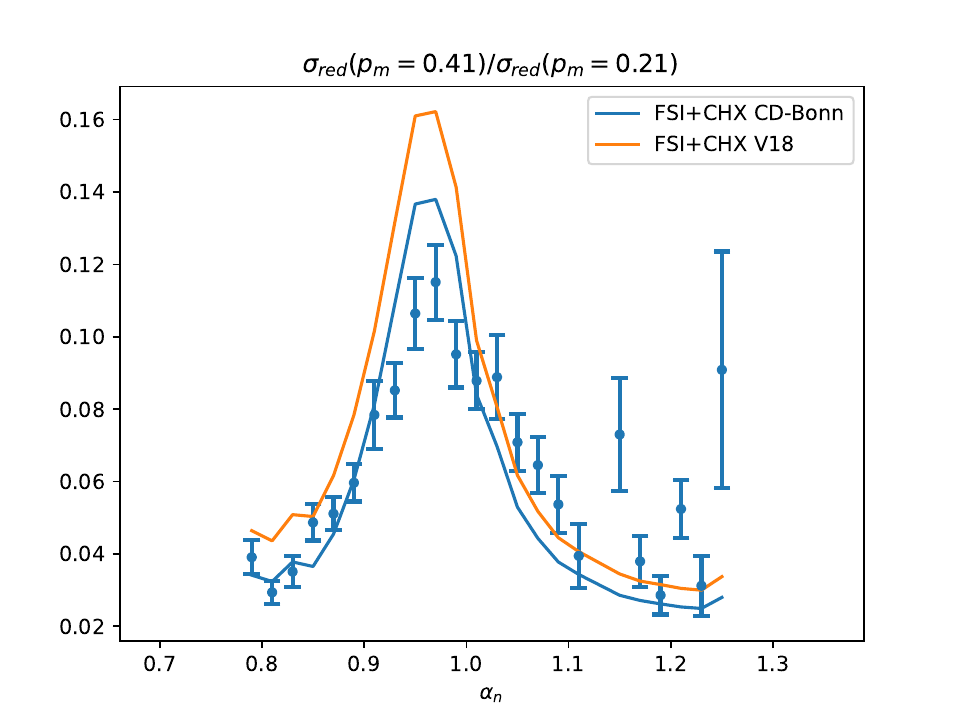}
\includegraphics[width=0.4\textwidth]{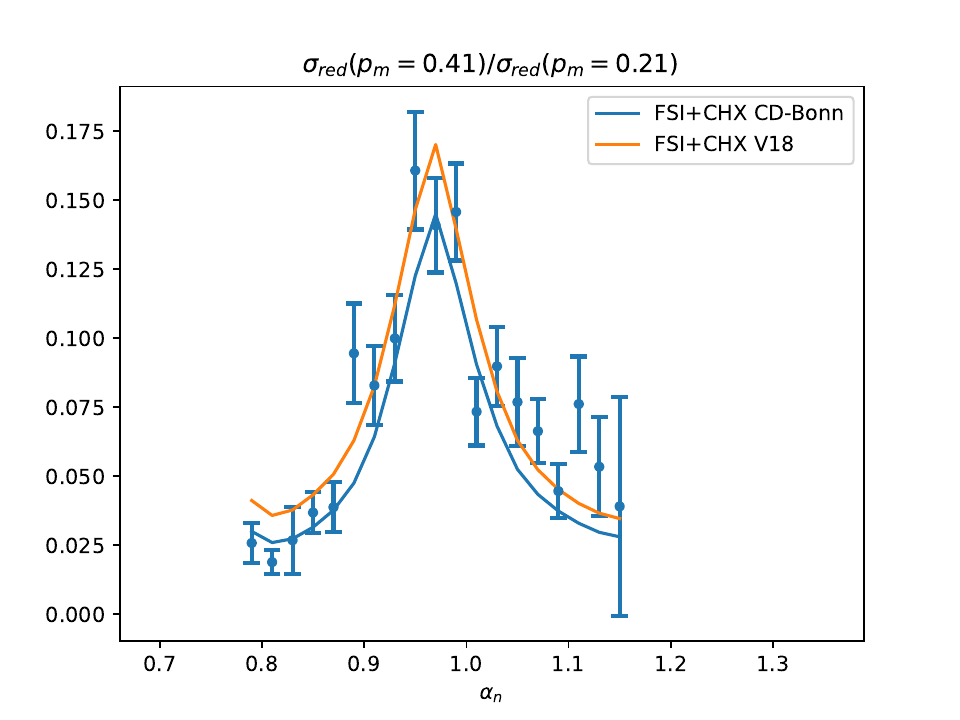}
\caption{\label{fig:dataDoubleRatio} 
On the left (right) is the reduced cross section ratio at $Q^2=2.1$~(GeV/$c)^2$ ($Q^2=3.5$~(GeV/$c)^2$) from the deuteron knockout experiment in~\cite{PhysRevLett.107.262501} as a function of $\alpha$~\cite{boeglin_private}. While there is some difference in the choice of $NN$ potential in describing the data, the difference is independent of $Q^2$. Furthermore, these data are in agreement with the calculations in Fig.~\ref{fig:rescatteringPrediction}.}
\end{figure}

Some detailed d$(e,e'p)n$ data exists from previous experiments at can already validate the predictions at lower $Q^2$. In Fig.~\ref{fig:dataDoubleRatio}, we see data from $Q^2=2.1$ and $3.5$~(GeV/$c)^2$) as a function of $\alpha$. The peak values of the cross section ratio are already in reasonable agreement with the ``no CT" prediction in Fig.~\ref{fig:rescatteringPrediction}. The prediction curves here include the calculations of FSI effects and single charge exchange (SCX) with different $NN$ potentials. The choice of $NN$ potential shows a systematic offset that is independent of $Q^2$ and, therefore, does not affect the observation of the CT signal which relies on the changes of the cross section ratio with respect to $Q^2$. The model predictions improve with increasing $Q^2$ because the accuracy of the generalized eikonal approximation improves at high momentum.  

\section{The Proposed Measurement}

\begin{figure}[htb]
\centering
\includegraphics[width=0.7\textwidth]{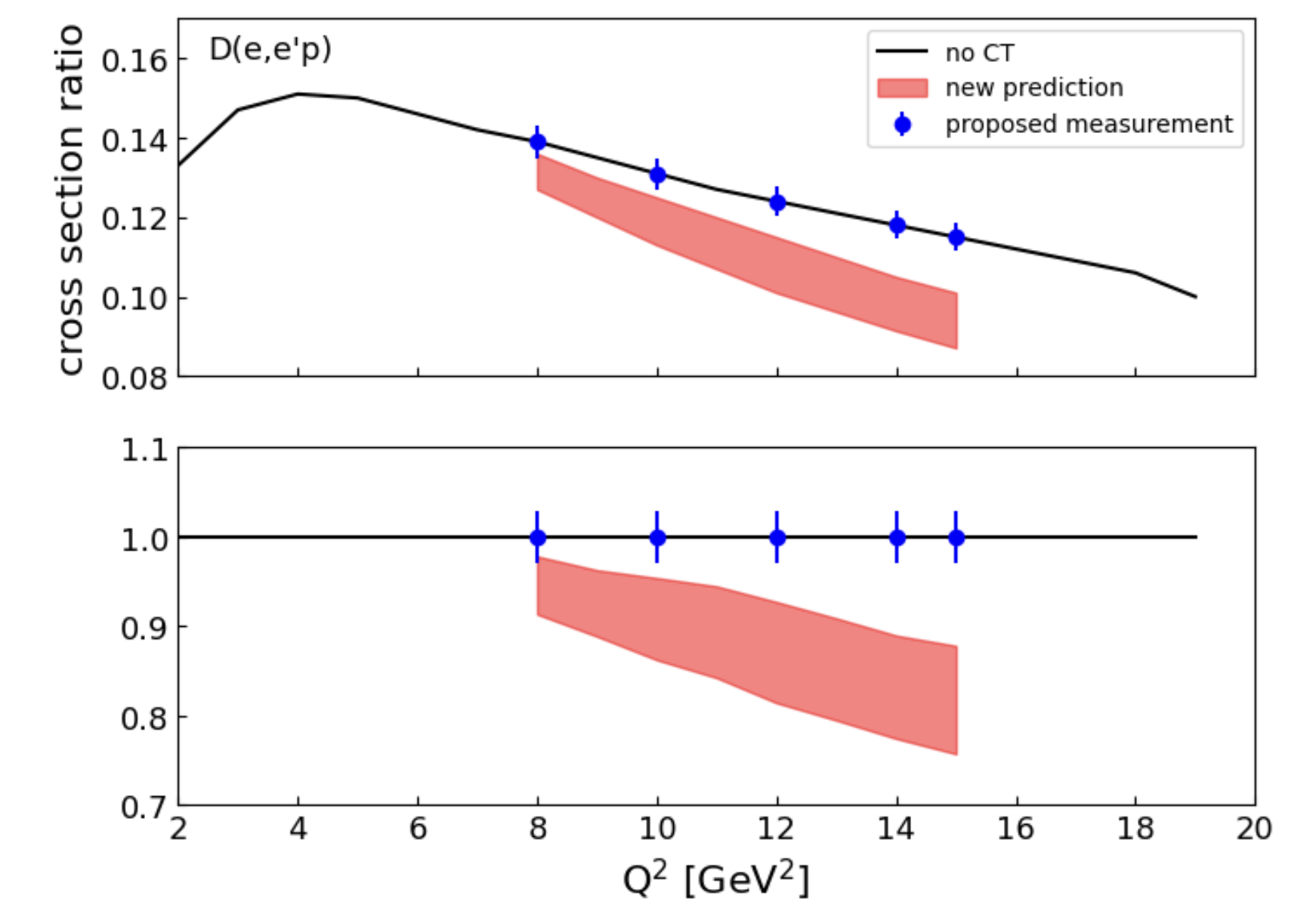}
\caption{\label{fig:expImpact} The cross section ratio $R=\frac{\sigma(p_r=400\textrm{MeV/c})}{\sigma(p_r=200\textrm{MeV/c})}$ in the d$(e,e'p)n$ reaction 
 (as from Fig.~\ref{fig:rescatteringPrediction}) overlaid with our proposed measurement. The shaded red band is the region between $\Delta M^2$=2-3~GeV$^2$. The blue points are the kinematics in this proposal with the the nominal 3.2\% statistical uncertainty.} 
\end{figure}

We propose to measure the onset of CT in d$(e,e'p)n$ by detecting the electron and proton in coincidence in Hall C from electron scattering on a deuterium target. We will construct ratios of the protons detected with high missing momenta ($300-600$ MeV$/c$) to those having low missing momenta ($50-150$ MeV$/c$) as a function of $Q^2$. We will explore a range of $Q^2$ that overlaps with the previous E12-06-107 experiment which was taken in parallel kinematics. By the time of this experimental running, we expect to receive 11~GeV of beam in Hall C. This, together with the full momentum capabilities of the Hall C spectrometers, will extend our measured $Q^2$ up to 15~(GeV$/c)^2$ as shown in Fig.~\ref{fig:expImpact}.

This measurement will increase our sensitivity for observing protons in a PLC. We will measure the ratio of experimental quantities with respect to experimental quantities, and we will compare this ratio with theory ratios taken with respect to theory. This is a different approach to the previous electron-scattering CT experiments which measure the ratios of experimental quantities direct with theory, and it reduces our overall systematic uncertainties. Uniquely, this experiment will measure proton knockout in a regime with high FSI contributions ($\alpha\approx1$). 

We plan to use a 25~cm liquid deuterium target(5$\%$ radiation length) to receive beam at 80~$\mu$A for a luminosity per proton of 6.23$\times10^{38}$~A$^{-1}$cm$^{-2}$s$^{-1}$. The 25~cm target will significantly increase our statistics compared to the standard Hall C 10~cm target, but it requires more cooling power, which will be possible after the ESR-2 upgrade. We use an aluminum dummy target for background target cell subtraction. The spectrometer detectors will be used in their standard configurations.

\subsection{Kinematics}
\label{ssec:kinematics}

\begin{figure}[htb!]
\centering
\includegraphics[width=1\textwidth]{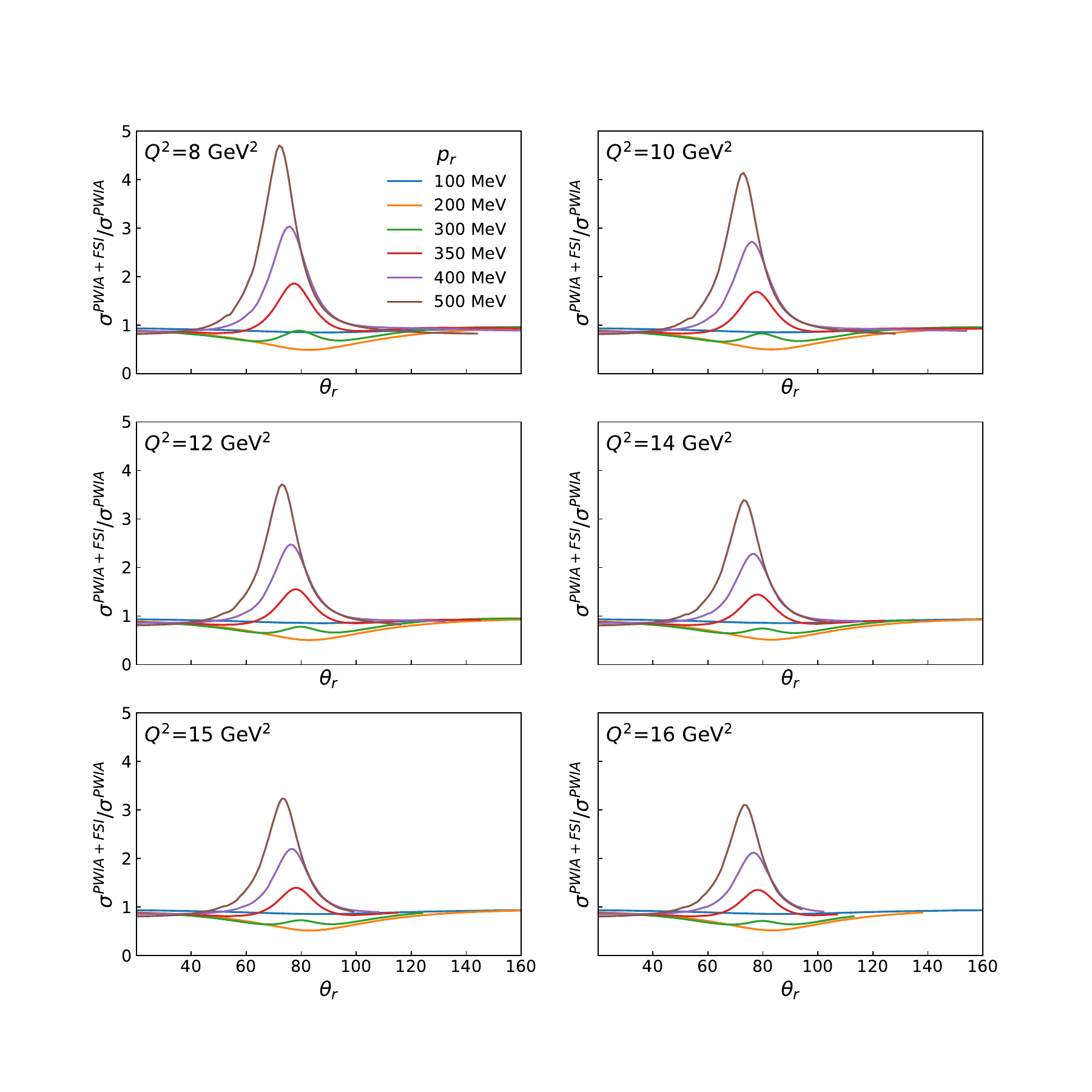}
\caption{\label{fig:rescatterCalc2} The dependence of the cross section ratio taken with respect to the PWIA on the recoil angle of the neutron with respect to $\vec{q}$ ($\theta_{rq}$) for different values of the recoil neutron momenta is shown for $Q^2=8$ to 16~(GeV$/c)^2$.} 
\end{figure}

While many of the previous electron scattering CT experiments on the proton have utilized parallel kinematics to reduce FSIs, here we employ perpendicular kinematics to increase FSIs and to measure CT in re-scattering reactions in an effort to improve our sensitivity to the formation of PLCs. We will use five kinematics settings to cover $Q^2$ from 8 to 15 GeV$^2$ (see Table~\ref{rates}). At each setting, the HMS will detect the scattered electrons at the quasi-elastic peak, while high-momentum struck protons will be measured in the SHMS. The SHMS momentum is centered at $p_\text{miss}=-200$ GeV$/c$ to cover both the low and high $p_\text{miss}$ regions simultaneously. We focus on the negative $p_\text{miss}$(backward kinematics where the struck proton goes backward with respect to $\vec{q}$) to get more counts and fewer background pions.
We calculated the ratio of (PWIA+FSI)/PWIA as a function of the recoiling neutron angle for each kinematic setting in Fig.~\ref{fig:rescatterCalc2}. The impact of the FSI is slightly reduced with increasing $Q^2$, but it still shows a strong enhancement at $p_\text{miss}$ between $300$ to $600$ MeV$/c$, and a suppression at $50 < p_\text{miss} < 150$ MeV$/c$. 

\begin{figure}[htb!]
\centering
\includegraphics[width=0.5\textwidth]{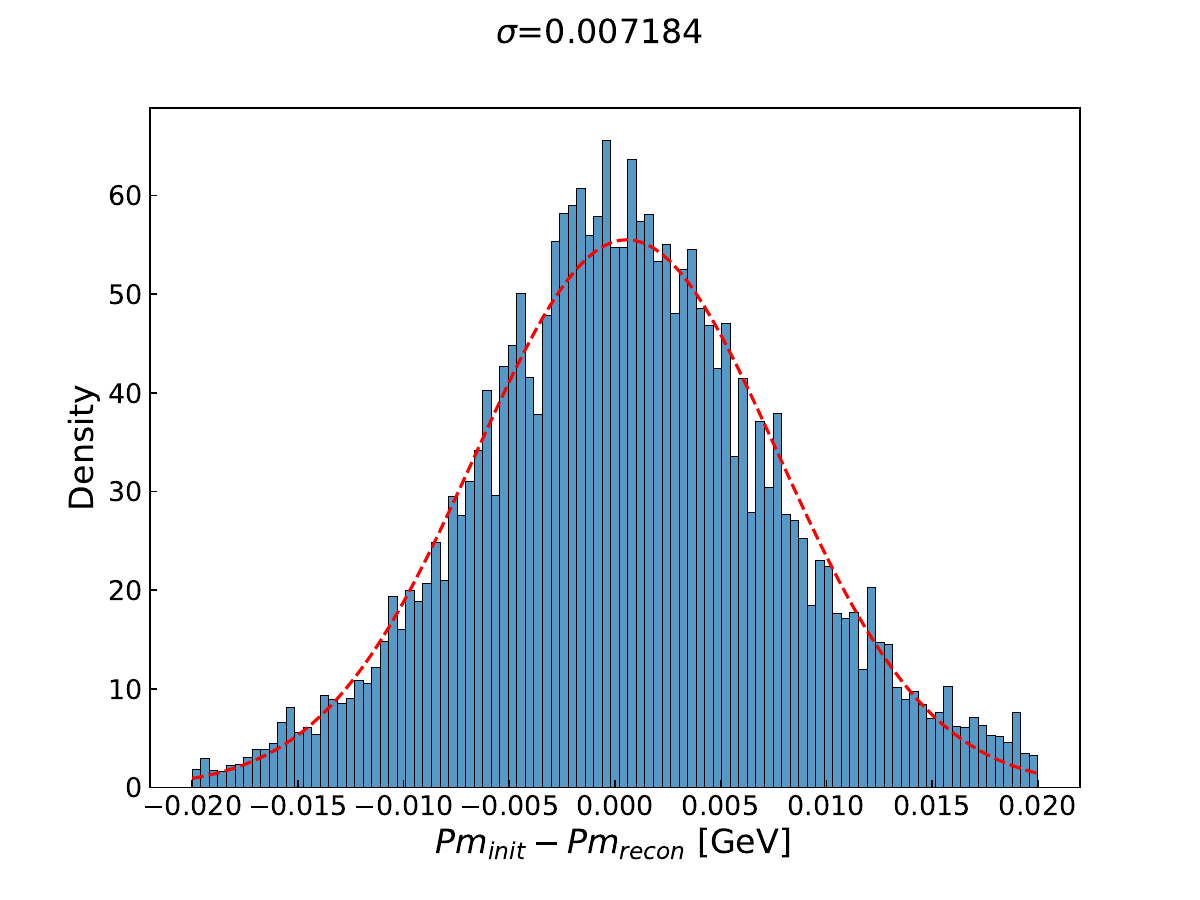}
\caption{\label{fig:pm_resol} The $P_m$ resolution at $Q^2=14$~(GeV$/c)^2$ is about 7 MeV.}
\end{figure}

\begin{figure}[htb!]
\centering
\includegraphics[width=1\textwidth]{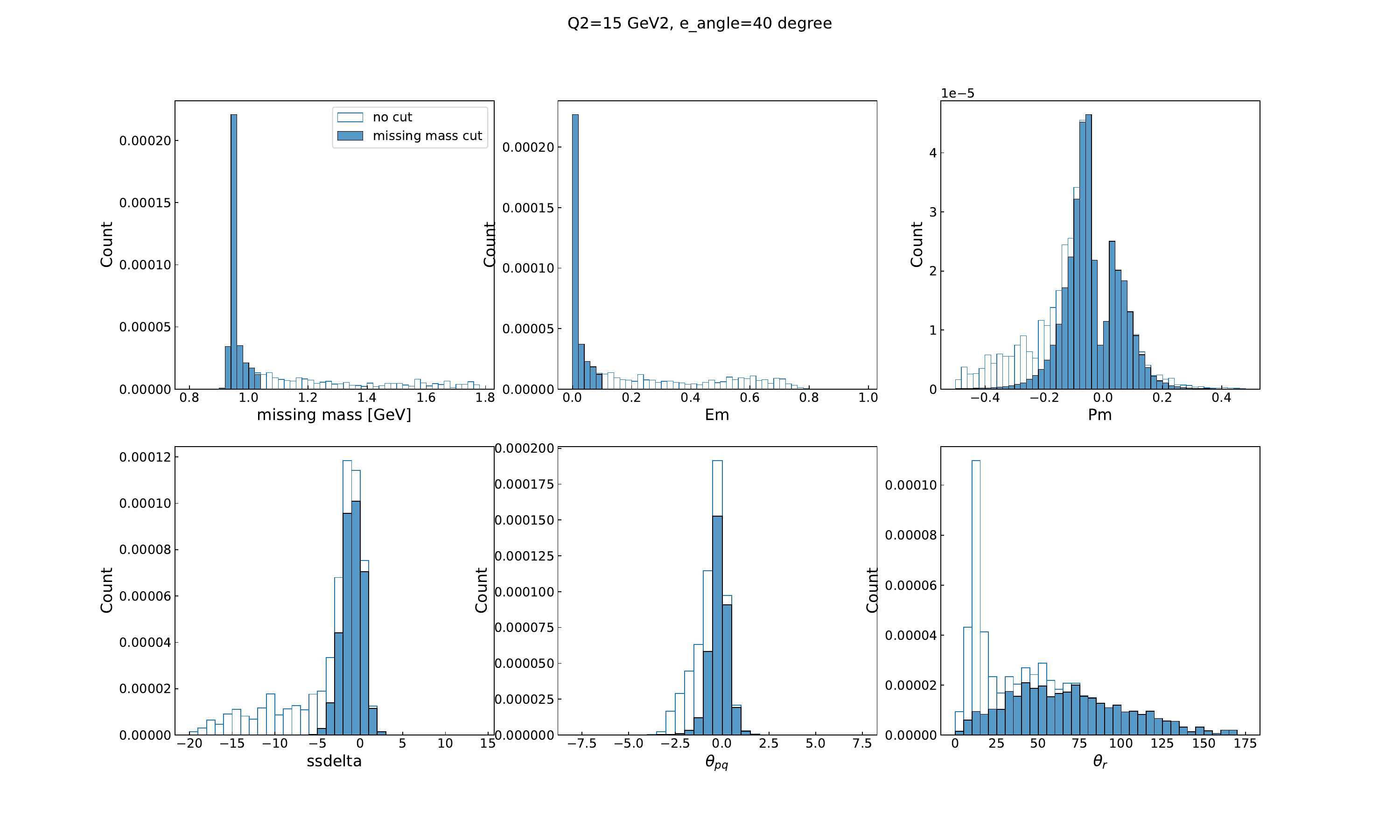}
\caption{\label{fig:kin_1d} Kinematics variables from SIMC before (non-filled) or after (filled, blue) a missing mass cut of neutron mass$\pm0.1$ GeV at $Q^2=15$~(GeV$/c)^2$.}
\end{figure}

The missing momentum resolution is shown in Fig.~\ref{fig:pm_resol} for the $Q^2=14$~(GeV$/c)^2$ setting and is shown to be approximately 7~MeV. This resolution is consistent with that observed in the previous CT Hall C measurement~\cite{PhysRevLett.126.082301}.
Furthermore, the missing mass is reconstructed at each kinematic setting. In our simulations, we reconstruct the missing mass of the neutron with and without radiative effects. We optimize our rate and kinematic studies with a cut on the neutron mass $\pm$0.1~GeV. 
\begin{figure}[htb!]
\centering
\includegraphics[width=0.5\textwidth]{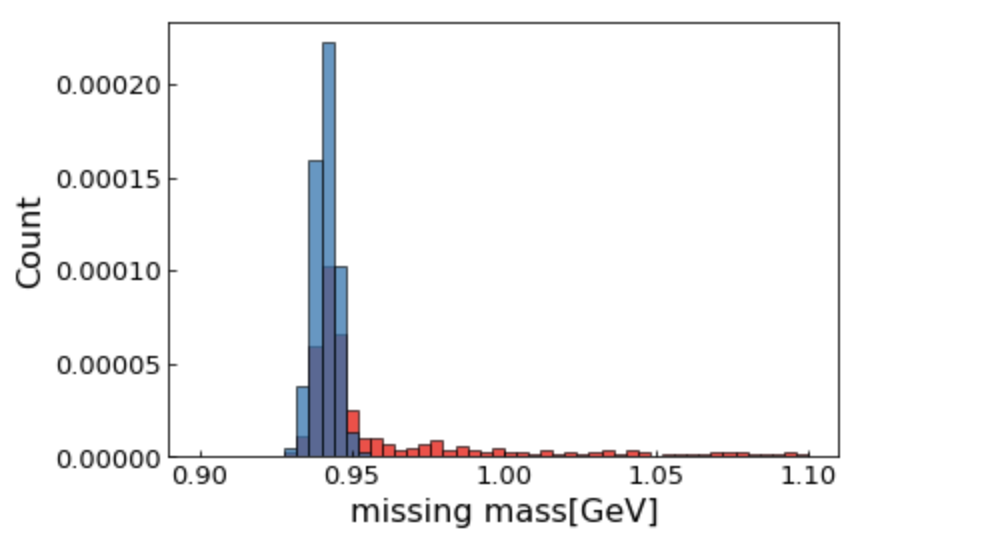}
\caption{\label{fig:missing_mass} The missing mass from SIMC with(red) and without(blue) radiative effect at $Q^2=14$~(GeV$/c)^2$ both peak at neutron mass.}
\end{figure}
The effect of the missing mass cut on the reconstructed neutron mass is shown in Fig.~\ref{fig:kin_1d}. This cut is essential to ensure that we have adequately reconstructed the recoiling nucleon. 

\begin{figure}[htb!]
\centering
\includegraphics[width=1\textwidth]{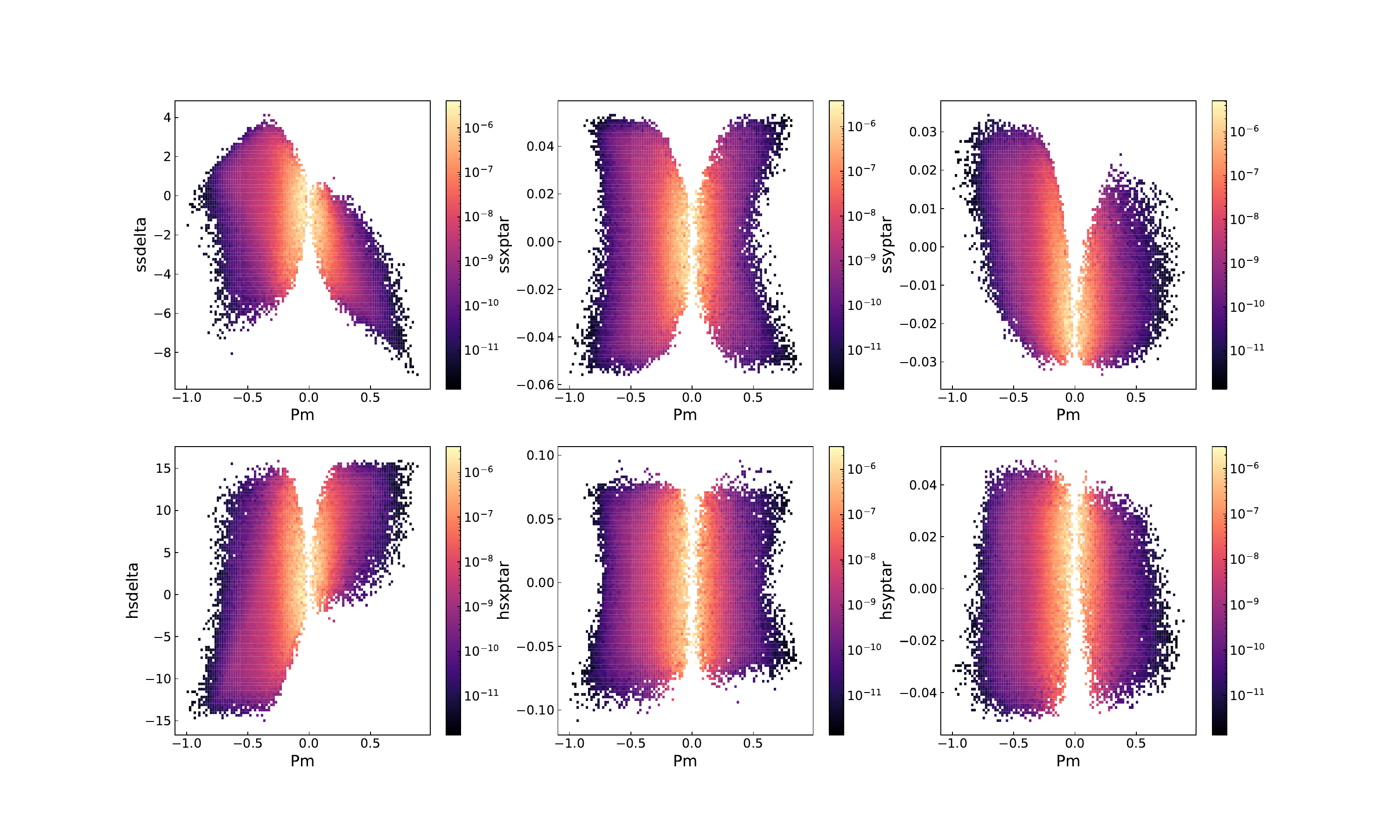}
\caption{\label{fig:pm_2d} SHMS and HMS target variables from SIMC with respect to $p_\text{miss}$ at $Q^2=14$~(GeV$/c)^2$ after the missing mass cut.}
\end{figure}


\begin{table*}[htb!]
\caption{\label{rates} Proposed Kinematics}
\centering
\begin{tabular}{cccccc}\hline 
Kinematics &$Q^2$ & $P_e$~(GeV/$c$) & $\theta_e$~(deg) & $P_p$~(GeV/$c$) & $\theta_p$~(deg) \\ 
\hline
1&8.046  & 6.713 & 19.000 & 5.121 & 27.380 \\
2&9.958  & 5.694 & 23.000 & 6.154 & 22.972 \\
3&11.941 & 4.637 & 28.000 & 7.222 & 19.073 \\
4&14.026 & 3.525 & 35.000 & 8.341 & 15.363 \\
5&15.127 & 2.939 & 40.000 & 8.931 & 13.461 \\
 \hline
\end{tabular}
\end{table*} 

\subsection{Expected Rates}
Kinematics and rate studies are done with the Hall C SIMC Monte-Carlo simulation package~\cite{simc}. The deuteron cross section is calculated with the CC1 off-shell prescription~\cite{DeForest:1983ahx} and the AV18 deuteron momentum distribution~\cite{PhysRevC.89.024305}. The effects of FSI are included by following the same calculation in \cite{Sargsian:2009hf}. No charge exchange are considered in the rate estimation. The expected coincidence rates per hour are integrated over two $p_\text{miss}$ region at each $Q^2$, see Table~\ref{total-rates}. The overall times for running on each target at each kinematic setting are shown for 1000 coincidence counts(corresponding to statistical uncertainty of approximately 3\%) in the high $p_\text{miss}$ region (kinematics b). Kinematical settings 1 and 2 are scaled to match statistics from the recent Hall C CT experiment, that is, 5k and 2k events respectively. An overall efficiency factor of 0.8 is applied to all time estimations and is estimated from the measured experimental inefficiency in the recent Hall C CT measurement.

\begin{table*}[htb!]
\caption{\label{total-rates} The expected $ep$ coincidence rates per hour are shown for each kinematics at two missing momentum ranges, a: 50-150 MeV, b: 300-600 MeV, along with their $P_m$,$\theta_r$,$Q^2$ in weighted average. ``PAC days" shows the time needed for each high $P_m$ region (b) to accumulate 1000 counts with an efficiency factor of 80$\%$ (except for settings 1 and 2 which will detect 5k and 2k events in this regime). SIMC with radiative effects included is used for this estimation, assuming 80 $\mu$A beam on a 25cm liquid deuterium target at 11 GeV.}
\centering
\begin{tabular}{ccccccc}\hline
\multicolumn{2}{c}{\textbf{Kinematics}} & \textbf{$P_m$} & \textbf{$\theta_r$} & \textbf{$Q^2$} & \textbf{Rate/hour} & \textbf{PAC days} \\\hline
\multirow{2}{*}{1}    &   a    &   0.08    &   79.06    &   7.49    &   5690.81    &   \multirow{2}{*}{1.5} \\
                      &   b    &   0.41    &   73.33    &   7.88    &   149.04    &   \\ \hline
\multirow{2}{*}{2}    &   a    &   0.08    &   77.15    &   9.52    &   1536.20    &   \multirow{2}{*}{3.0} \\
                      &   b    &   0.41    &   74.44    &   9.77    &   36.47    &   \\ \hline
\multirow{2}{*}{3}    &   a    &   0.08    &   77.40    &   11.62    &   413.28    &   \multirow{2}{*}{5.7} \\
                      &   b    &   0.41    &   75.46    &   11.70    &   9.16    &   \\ \hline
\multirow{2}{*}{4}    &   a    &   0.09    &   77.28    &   13.76    &   93.63    &   \multirow{2}{*}{25.1} \\
                      &   b    &   0.40    &   75.95    &   13.74    &   2.07    &   \\ \hline
\multirow{2}{*}{5}    &   a    &   0.09    &   78.73    &   14.94    &   36.63    &   \multirow{2}{*}{55.9} \\
                      &   b    &   0.40    &   75.72    &   14.83    &   0.93    &   \\\hline
\end{tabular}
\end{table*} 
The total measurement times shown in Table~\ref{total-rates} are given in PAC days. Additional running on the aluminum dummy cell for the window background subtraction will require approximately two percent of the beam time used for the full target (adding an additional 2 days of beam in total). Configuration changes and calibration runs are expected to take 2 more PAC days. The total time requested for this experiment is approximately 95~days.  

\subsection{Systematics}

We assume the measured systematics from the previous Hall C measurement of 4\% as an estimate of what to expect in this experiment. We will use ratios which will reduce some of the magnitude of the systematics. The largest systematics from the Hall C measurement were on the spectrometer acceptance (2.6\%) which we expect to reduce from better knowledge of the spectrometers and with the use of experimental transparency ratios taken at the same kinematics. The next largest sources of uncertainty were on the knowledge of the fundamental $ep$ elastic cross section (1.8\%) and the proton absorption in materials between the target and detector (1.2\%). Both of these effects will minimized as systematic contributions through the use of experimental ratios. We note that while the specific choice of NN potential has some difference, the difference is independent of $Q^2$ and does not reduce the sensitivity to the signal extraction (see Fig.~\ref{fig:dataDoubleRatio}). 


\section{Complementary Experiments at JLab}

Experiments in the JLab 6~GeV era observed the onset of CT in mesons. Hall C observed the onset of CT in pion electroproduction in E01-107 in the range of $Q^2=1$-5~GeV/c$^2$~\cite{Clasie:2007aa}~\cite{Qian:2009aa} on $^{12}$C, $^{27}$Al, $^{63}$Cu, and $^{197}$Au targets. CLAS experiment E02-110 measured the onset in rho electroproduction on $^{12}$C and $^{56}$Fe targets at $Q^2$ between 0.8-2.4~GeV/c$^2$~\cite{ElFassi:2012nr}. The CLAS experiment E12-06-106~\cite{E12:06:106} further investigates the nuclear target and $Q^2$ dependence of CT in rho meson electroproduction up to $Q^2=5.5$~GeV/c$^2$. The onset of CT in mesons confirms the presence of hadrons in point-like configurations (PLCs) and further motivates the search for baryons in such configurations. The $Q^2$ at which CT begins in mesons versus barons is unknown as well as the nucleus $A$-dependence, but the measurement of the onset in baryons will elucidate information about PLC expansion times and effects in nuclear physics. 

After the 12~GeV upgrade, E12-06-107~\cite{E12:06:107} ruled out the onset of CT in $A(e,e'p)$ reactions up to $Q^2=14.2$~GeV/c$^2$ in parallel kinematics. No other experiment will have measured the CT effect in our proposed kinematics, but it should be noted that this d$(e,e'p)$ strategy has been a key approach to observing CT that was even included in motivation for the 12~GeV upgrade~\cite{Sargsian.2003}. The onset of CT is a requirement for the validity of the QCD factorization theorem ~\cite{Strikman:2000qn} for exclusive meson production in DIS (a critical experimental and theoretical effort at JLab). 

We propose 95~days of beam running in Hall C with a 11~GeV electron beam at 80~$\mu$A to measure the d$(e,e'p)$ reaction in the standard HMS and SHMS spectrometers in coincidence. This measurement will extend the previous measurements that did not observe the onset of CT in protons in parallel kinematics, but we will search for the onset of CT in transverse kinematics characterized as having higher FSIs and with an increased sensitivity and a clear separation between the production mechanism and expansion of a PLC. 


\newpage
\bibliographystyle{ieeetr}
\bibliography{refs}

\end{document}